\documentclass[11pt,notitlepage]{article}

\usepackage[scaled=0.97]{helvet}
\usepackage[T1]{fontenc}

\usepackage{array}
\usepackage{amsmath,amsbsy,amsthm}
\usepackage{amsxtra}
\usepackage{amstext}
\usepackage{amssymb}
\usepackage{latexsym}

\usepackage{url}
\usepackage{graphicx}

\usepackage[usenames]{color}   

\usepackage{setspace}          
\usepackage[compact]{titlesec} 

\usepackage{natbib}

\usepackage{adjustbox,booktabs}
\usepackage[top=1in,left=1in,right=1in,bottom=1in]{geometry}

\usepackage{titlesec}          
\usepackage{enumitem}          
\usepackage{wrapfig}           

\newcommand{\bA}{\boldsymbol{A}}

\newcommand{\bH}{\boldsymbol{H}}
\newcommand{\bI}{\boldsymbol{I}}

\newcommand{\bY}{\boldsymbol{Y}}

\newcommand{\bbeta}{\boldsymbol{\beta}}

\newcommand{\bnu}{\boldsymbol{\nu}}

\newcommand{\bDelta}{\boldsymbol{\Delta}}

\newcommand{\bSigma}{\boldsymbol{\Sigma}}

\newcommand{\bPhi}{\boldsymbol{\Phi}}

\newcommand{\mendel}{{\sc Mendel}}
\newcommand{\Mendel}{{\sc Mendel}}

\title{Fast Genome-Wide QTL Analysis Using \Mendel}
\date{}
\author{
Hua Zhou \\
Department of Statistics \\
North Carolina State University \\
Raleigh, NC 27695-8203 \\
Email: hua\_zhou@ncsu.edu
\and
Jin Zhou \\
Department of Epidemiology and Biostatistics  \\
University of Arizona \\
Tucson, AZ 85721-0066 \\
Email: jzhou@email.arizona.edu
\and
Tao Hu \\
Bioinformatics Research Center \\
North Carolina State University \\
Raleigh, NC 27695 \\
Email: thu3@ncsu.edu
\and
Eric M. Sobel \\
Department of Human Genetics \\
University of California \\
Los Angeles, CA 90095-1766 \\
Email: esobel@mednet.ucla.edu
\and
Kenneth Lange \\
Departments of Biomathematics, \\
Human Genetics, and Statistics \\
University of California \\
Los Angeles, CA 90095-1766 \\
Email: klange@ucla.edu
}

\begin{document}

\baselineskip=10pt
\maketitle

\onehalfspacing

\pagestyle{empty}

\begin{abstract}
\noindent
Pedigree GWAS (Option 29) in the current version of the \mendel\ software is an optimized subroutine for performing large scale genome-wide QTL analysis. This analysis (a) works for random sample data, pedigree data, or a mix of both, (b) is highly efficient in both run time and memory requirement, (c) accommodates both univariate and multivariate traits, (d) works for autosomal and x-linked loci, (e) correctly deals with missing data in traits, covariates, and genotypes, (f) allows for covariate adjustment and constraints among parameters, (g) uses either theoretical or SNP-based empirical kinship matrix for additive polygenic effects, (h) allows extra variance components such as dominant polygenic effects and household effects, (i) detects and reports outlier individuals and pedigrees, and (j) allows for robust estimation via the $t$-distribution. The current paper assesses these capabilities on the genetics analysis workshop 19 (GAW19) sequencing data. We analyzed simulated and real phenotypes for both family and random sample data sets. For instance, when jointly testing the 8 longitudinally measured systolic blood pressure (SBP) and diastolic blood pressure (DBP) traits, it takes \mendel\ 78 minutes on a standard laptop computer to read, quality check, and analyze a data set with 849 individuals and 8.3 million SNPs. Genome-wide eQTL analysis of 20,643 expression traits on 641 individuals with 8.3 million SNPs takes 30 hours using 20 parallel runs on a cluster. \Mendel\ is freely available at \url{http://www.genetics.ucla.edu/software}.

\end{abstract}

\section{Background}

The classical variance component model has been a powerful tool for mapping quantitative trait loci (QTL) in pedigrees.  Polygenic effects are effectively modeled by introducing an additive genetic variance component operating on the kinship coefficient matrix. With unknown or dubious pedigree structure, global kinship coefficients can be accurately estimated from dense markers using either the genetic relationship matrix (GRM) or the method of moments. In GWAS (genome-wide association studies), the two alleles of a SNP (single nucleotide polymorphism) shift trait means and can be tested as a fixed effect. 
However, fitting a variance component model is computationally challenging, especially when it has to be done for a large number of markers. In the newly released version of the \mendel\ software \citep{Lange13Mendel}, Option 29 implements an ultra-fast score test for pedigree GWAS. Score tests require no additional iteration under the alternative model.  Only SNPs with the most promising score-test p-values are further subject to likelihood ratio testing (LRT), thus achieving a good compromise between speed and power for large scale QTL analysis. 
In this paper, we demonstrate the capabilities of \mendel\ on the genetic analysis workshop 19 (GAW19) sequencing data.

\section{Methods}


QTL association mapping typically invokes the multivariate normal distribution to model the observed $T$-variate trait $\bY \in \mathbb{R}^{n \times T}$ over a pedigree of $n$ individuals. The standard model \citep{Lange02GeneticsBook} collects the means of the responses $\mathrm{vec}(\bY)$ into a vector $\bnu$ and the corresponding covariances into a matrix $\bSigma$ and represents the loglikelihood of a pedigree as
\begin{eqnarray*}
L & = & -\frac{1}{2} \ln \det \bSigma
        -\frac{1}{2}[\mathrm{vec}(\bY)-\bnu]^t \bSigma^{-1}[\mathrm{vec}(\bY)-\bnu],
\end{eqnarray*}
where the covariance matrix is typically parametrized as $\bSigma = 2 \bSigma_a \otimes \bPhi + \bSigma_d \otimes \bDelta_7 + \bSigma_h \otimes \bH + \bSigma_e \otimes \bI$.
Here $\bPhi$ is the global kinship matrix capturing additive polygenic effects, and $\bDelta_7$ is a condensed identity coefficient matrix capturing dominance genetic effects. For $\bPhi$, \mendel\ can use (a) the theoretical kinship matrix from provided pedigree structures, (b) SNP-based estimates for the kinship of pairs of people within each pedigree, or (c) SNP-based estimates for the entire global kinship matrix ignoring pedigree information. To estimate kinship coefficients from dense SNP data, \mendel\ employs either the genetic relationship matrix (GRM) or the method of moments \citep{Day-Williams11LinkageWoPedigree,Lange14NextGenStatGene}. The household effect matrix $\bH$ has entries $h_{ij}=1$ if individuals $i$ and $j$ are in the same household and 0 otherwise. Individual environmental contributions and trait measurement errors are incorporated via the identity matrix $\bI$. 
QTL fixed effects are captured through the mean component $\bnu = \bA \bbeta$ for some predictor matrix $\bA$ and vector of regression coefficients $\bbeta$.  To test a SNP against a $T$-variate trait, $\bA$ is augmented with $T$ extra columns holding the allele counts at the SNP, and the corresponding regression coefficients are jointly tested for association \citep{Lange05association}.
For longitudinal measurements of covariates such as {\tt smoke}, {\tt age} and {\tt BPMed}, we may either assume time varying effect sizes or constrain their effect sizes at different time points to be the same. The latter tactic leads to a more parsimonious and interpretable model and can be easily enforced by setting appropriate parameter constraints in \mendel's control file, which lists the user's choice of model parameters. In \mendel, SNPs with the most impressive score test p-values (top 10 by default) are further tested by the more accurate, but slower, likelihood ratio method, thus achieving a good compromise between speed and power for large-scale QTL analysis. We refer readers to our companion manuscript \citep{Zhou13PedScore} for more model and implementation details.

\section{Results for Family Data}


Our analyses are based on the genotype calls for 959 individuals (464 directly sequenced and the rest imputed) provided in the {\tt chrX-geno.csv.gz} files. 
Section~\ref{sec:size-power} is a size and power study of the simulated traits in all 200 replicates (SIMPHEN.1-200).
Section~\ref{sec:real-dbp} is the whole genome QTL analysis for the real, systolic (SBP) and diastolic (DBP) blood pressure traits.

\subsection{Size and power study using simulated traits (SIMPHEN.1--200)}
\label{sec:size-power}

The power to detect the six functional variants in the {\tt MAP4} gene on chromosome 3 are evaluated from the 200 simulation replicates of the trivariate traits SBP and DBP. Type I errors are evaluated from the univariate Q1 trait, which does not involve a major gene. Our analysis includes covariates {\tt sex}, {\tt age}, {\tt BPMed}, {\tt Smoke}, and their pairwise interactions, and uses the theoretical kinship matrix as the additive polygenic variance component. We constrain the covariate effects to be equal across 3 time points. Table~\ref{table:map4-pvalues} shows that the type I error is well controlled. Not surprisingly the power for detecting the two rare functional variants 3-47913455 and 3-47957741 is extremely low.

\begin{table}
\centering
{\footnotesize
\begin{tabular}{ccccccrccccc}
\toprule
& & \multicolumn{3}{c}{($\text{DBP}_1$, $\text{DBP}_2$,$\text{DBP}_3$)} & \phantom & \multicolumn{3}{c}{$(\text{SBP}_1,\text{SBP}_2,\text{SBP}_3)$} & \phantom & \multicolumn{1}{c}{Q1} \\
\cmidrule{3-5}  \cmidrule{7-9} \cmidrule{11-11}
SNP & MAF & $\beta_{\text{DBP}}$ & \%Var & Power & \phantom & $\beta_{\text{SBP}}$ & \%Var & Power & \phantom & Size \\
\midrule
3-47913455 & 0.0049 & -5.4633 & 0.0036 & $0.05 \pm 0.02$ & \phantom &  -8.7001 & 0.0044 & $0.06 \pm 0.02$ & \phantom & $0.06 \pm 0.02$ \\
3-47956424 & 0.3777 & -1.4951 & 0.0117 & $0.35 \pm 0.03$ & \phantom &  -2.3810 & 0.0143 & $0.42 \pm 0.03$ & \phantom & $0.03 \pm 0.01$ \\
3-47957741 & 0.0016 & -5.0841 & 0.0024 & $0.04 \pm 0.01$ & \phantom &  -8.0964 & 0.0030 & $0.06 \pm 0.02$ & \phantom & $0.06 \pm 0.02$ \\
3-47957996 & 0.0301 & -4.6435 & 0.0122 & $0.82 \pm 0.03$ & \phantom &  -7.3946 & 0.0149 & $0.89 \pm 0.02$ & \phantom & $0.05 \pm 0.01$ \\
3-48040283 & 0.0318 & -6.2235 & 0.0229 & $0.84 \pm 0.03$ & \phantom &  -9.9107 & 0.0278 & $0.89 \pm 0.02$ & \phantom & $0.05 \pm 0.01$ \\
3-48040284 & 0.0131 & -6.9531 & 0.0091 & $0.47 \pm 0.04$ & \phantom & -11.0726 & 0.0111 & $0.56 \pm 0.06$ & \phantom & $0.04 \pm 0.01$ \\
\bottomrule
\end{tabular}
}
\caption{Empirical power for testing trivariate DBP and SBP traits and empirical type I error for testing the univariate Q1, based on simulation data in files SIMPHEN.1--SIMPHEN.200.}
\label{table:map4-pvalues}
\end{table}

\subsection{QTL analysis of the real, 8-variate phenotype ($\text{DBP}_i, \text{SBP}_i, i=1,2,3,4$)}
\label{sec:real-dbp}

SBPs and DBPs measured at 4 time points are available for 1389 members from 20 extended families. The largest family contains 107 individuals; the smallest, 27. Genotypes at 8,348,674 SNPs were available on 959 of the individuals. We analyzed all SNPs and pedigrees together for the 8-variate trait ($\text{SBP}_i,\text{DBP}_i,i=1,2,3,4$). Our model includes covariates {\tt sex}, {\tt age}, {\tt BPMed}, {\tt Smoke}, and their pairwise interactions and we constrain the covariate effects to be equal across 4 time points. The log-likelihoods of the null model (no SNPs included) using the theoretical kinship, GRM within pedigrees, or GRM across all individuals are -11675.95, -11696.90, and -11698.71 respectively, indicating that the provided pedigree information captures additive genetic effects adequately. The results summarized below use the theoretical kinship matrix.

To read in all the data and run standard quality control (QC) procedures took just under 5 minutes. QC excluded 10,603 SNPs and 110 individuals based on genotyping success rates below 98\%. The remaining 8,338,071 SNPs and 849 individuals are analyzed. The subsequent ped-GWAS analysis ran in 73 minutes for all results reported in Table~\ref{table:realpheno-DBP-SBP}.  Because we excluded rare SNPs with low minor allele frequencies ($< 0.03$) across 849 individuals, p-values were calculated for only 3,084,046 SNPs. Accordingly the genome-wide significance threshold is $1.62 \times 10^{-8}$ or 7.79 on the $\log_{10}$ scale; the threshold for a false discovery rate (FDR) of 0.05 is $4.19 \times 10^{-8}$ or 7.38 on the $\log_{10}$ scale.

Estimates for environmental effects and their interactions under the null model (no SNPs included) are listed in the top panel of Table~\ref{table:realpheno-DBP-SBP}. The middle panel displays the Manhattan and QQ plots. The genomic inflation factor of 1.023 indicates no systematic bias. One SNP passes the Bonferroni corrected genome-wide significance level, and three SNPs pass the FDR 0.05 threshold. They are listed in the bottom panel of Table~\ref{table:realpheno-DBP-SBP}. SNP 1-142617328 has a Hardy-Weinberg equilibrium (in founders) p-value less than $10^{-22}$, indicating possible genotyping error. The remaining two significant SNPs occur at 118,783,424 and 118,767,564 base pairs, respectively, on chromosome 11. Both show a MAF of 0.02778 in 413 founders. Since their MAFs in all 849 individuals are higher than 0.03, they were not removed in the filtering stage.

\begin{table}[th!]
\centering
{\footnotesize
  \begin{tabular}{@{} lrr @{}}
    \toprule
Mean effects & $\text{SBP}_1,\text{SBP}_2,\text{SBP}_3,\text{SBP}_4,$ & $\text{DBP}_1,\text{DBP}_2,\text{DBP}_3,\text{DBP}_4$ \\
\midrule
$\beta_{\text{Sex}}$ & 10.21 & 4.24 \\ 
$\beta_{\text{Age}_i}$ & 0.32 & 0.02 \\ 
$\beta_{\text{BPMed}_i}$ & 3.11 & 10.07 \\ 
$\beta_{\text{Smoke}_i}$  & 1.53 & 1.84 \\ 
$\beta_{\text{Sex}_i \times \text{BPMed}_i}$ & -3.20 & -1.84 \\ 
$\beta_{\text{Sex}_i \times \text{Smoke}_i}$ & 0.30 & -0.73 \\ 
$\beta_{\text{Sex}_i \times \text{Age}_i}$ & 0.41 & 0.14 \\ 
$\beta_{\text{BPMed}_i \times \text{Smoke}_i}$ & 3.83 & 2.43 \\ 
$\beta_{\text{BPMed}_i \times \text{Age}_i}$ & 0.01 & -0.35 \\ 
$\beta_{\text{Smoke}_i \times \text{Age}_i}$ & -0.06 & -0.06 \\ 
\bottomrule
  \end{tabular}
}
$$
\begin{array}{m{3.25in}m{3.25in}}
\includegraphics[width=3.25in]{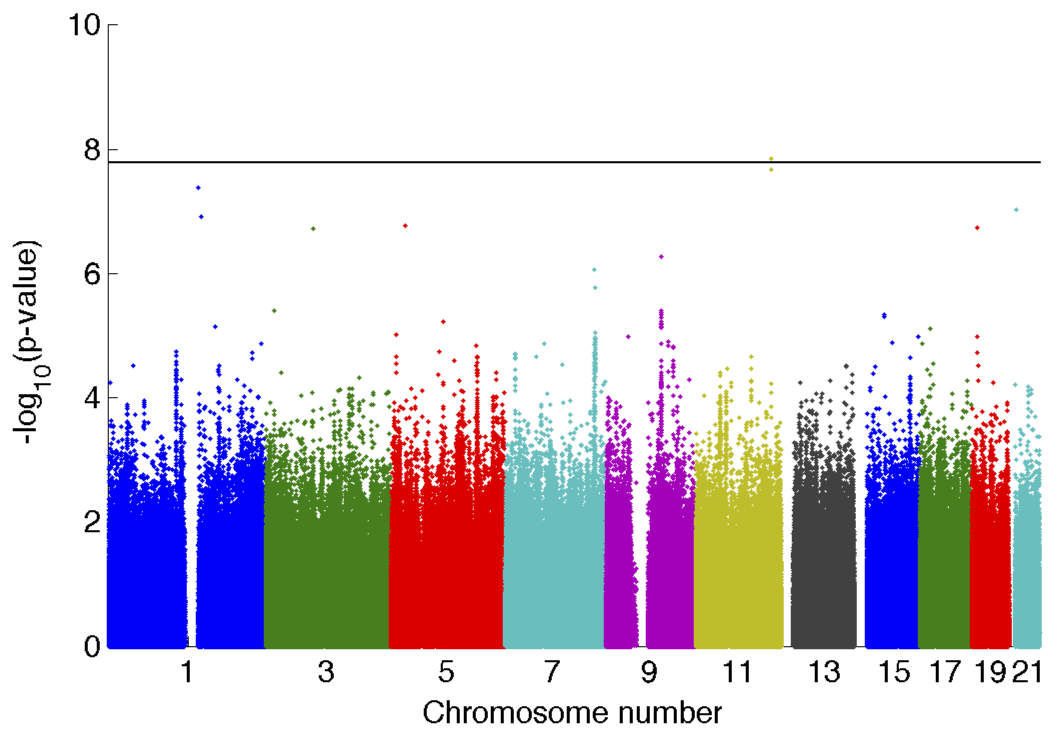} & \includegraphics[width=2.75in]{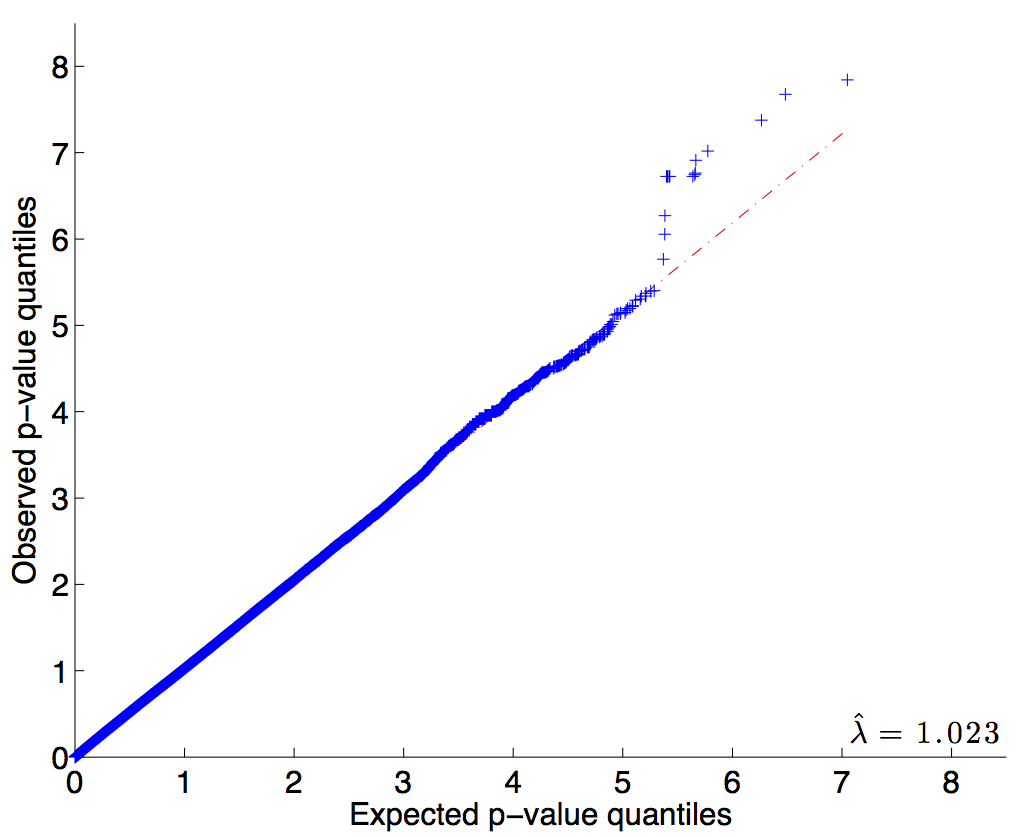}
\end{array}
$$
{\footnotesize
\begin{tabular}{rrrccc}
\toprule
SNP          & Chr. &   Base Pair  & MAF in founders & $-\log_{10}(\text{p-value})$ & HW p-value \\
\midrule
11-118783424 &  11  &  118,783,424 &        0.02778  &  7.84                        & 0.7665     \\
11-118767564 &  11  &  118,767,564 &        0.02778  &  7.68                        & 0.7665     \\
 1-142617328 &   1  &  142,617,328 &        0.49074  &  7.38                        & 0.0000     \\
\bottomrule
\end{tabular}
}
\caption{Multivariate QTL analysis of the real, 8-variate trait ($\text{SBP}_i,\text{DBP}_i, i=1,2,3,4$) from the family data with 849 individuals and 3.1 million SNPs (after filtering). Top: estimated mean effects under the null model (no SNPs included) using the theoretical kinship matrix for the additive polygenic variance component. Middle: Manhattan plot (left) and QQ plot (right). The horizontal line represents the genome-wide significance level. Bottom: Three SNPs that pass the FDR 0.05 threshold. The top SNP, 11-118783424, also passes the genome-wide significance level. The total run time on a laptop with an Intel Core i7 2.6 GHz CPU and 16 GB RAM is {\bf 78 minutes}.}
\label{table:realpheno-DBP-SBP}
\end{table}


\subsection{Genome-wide eQTL analysis of 20,634 expression traits}

Genome-wide measures of 20,634 gene expression levels in peripheral blood mononuclear cells (PBMCs) from the first study examination are provided for 643 individuals in the family data. The formidable task of exhaustive eQTL analysis (20,634 expressions vs 8,338,071 SNPs) can be easily managed using \mendel. We submitted 20 parallel jobs to a cluster and finished the complete analysis in about 30 hours. 

\begin{figure}[t]
\centering
$$
\begin{array}{cc}
\includegraphics[width=3in]{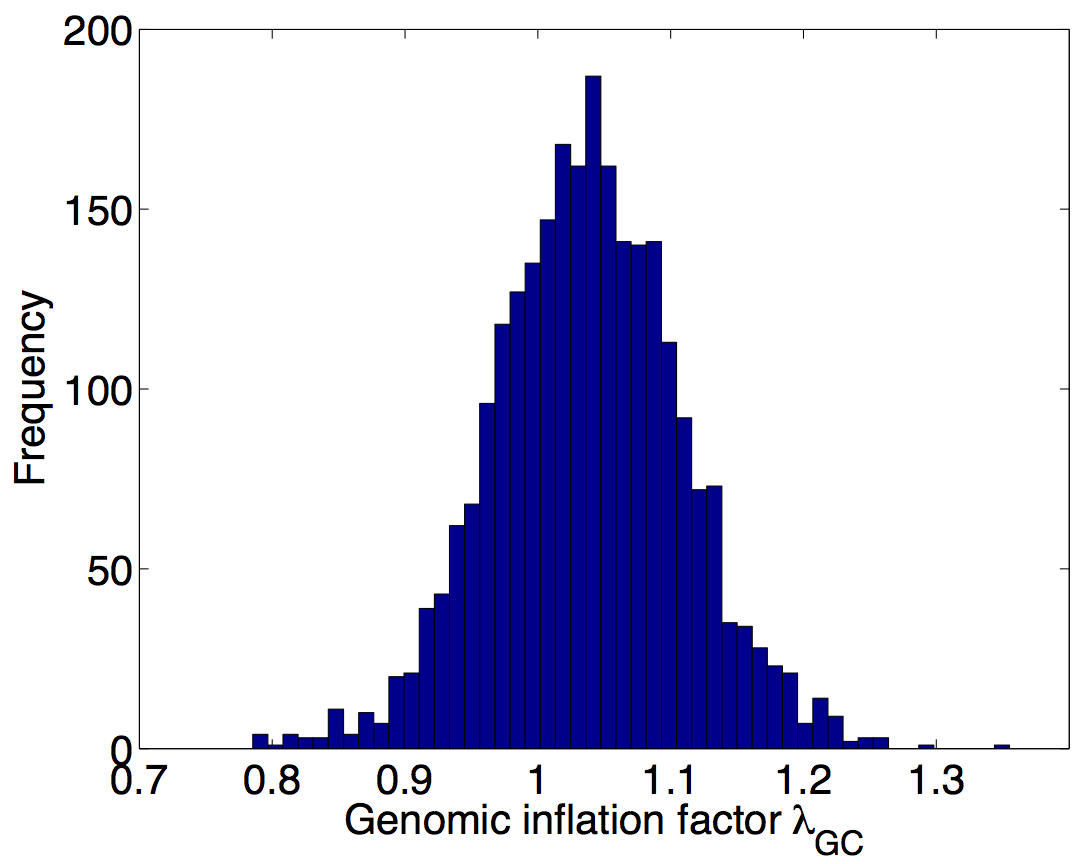} & \includegraphics[width=3in]{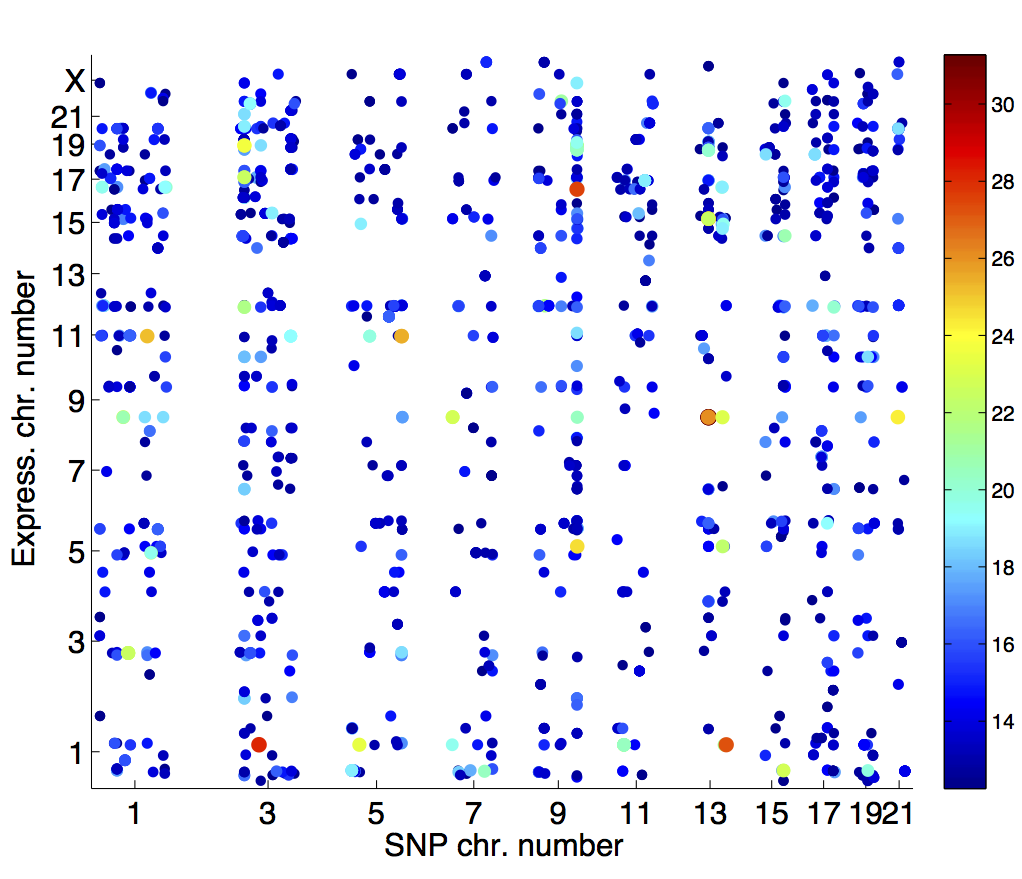}
\end{array}
$$
\caption{Summary of the eQTL analysis. Left: Histogram of the genomic inflation factors $\lambda_{GC}$. Right: Top expression-SNP hits from the eQTL analysis. Each dot represents an expression-SNP association that satisfies: genomic inflation factor $\lambda_{GC}<1.1$, $\text{p-value}<5.77 \times 10^{-13}$, SNP Hardy-Weinberg test (in founders) p-value $>10^{-8}$, SNP MAF in 641 individuals $>0.01$, and the expression probe is annotated in the {\tt  EXPR\_MAP.csv} file. Dot size and color vary according to their p-values on the $\log_{10}$ scale. Total run time (20,634 expressions vs 8,338,071 SNPs) on a cluster with 20 parallel jobs is about {\bf 30 hours}.}
\label{fig:eQTL-summary}
\end{figure}

In all eQTL runs, SNPs and individuals with genotyping success rate $\le 0.98$ are excluded from analysis. Rare variants with MAF $\le 0.01$ in all individuals are also excluded. This leaves 641 individuals and 4,199,714 SNPs. The theoretical kinship matrix is used for the additive polygenic variance component. Our analysis includes covariates {\tt sex}, {\tt age}, {\tt BPMed}, {\tt Smoke}, and their pairwise interactions. Initialization takes about 5 minutes; the subsequent genome-wide QTL mapping of each expression trait takes about 1 to 2 minutes. The left panel of Figure~\ref{fig:eQTL-summary} displays a histogram of genomic inflation factors from 20,634 genome-wide QTL analyses. They are well-concentrated around 1 and indicate no or little systematic bias. The right panel shows the top hits that satisfy a set of stringent criteria listed in the figure caption. Note that the whole eQTL significance level is set at $0.05/20634/4199714=5.77 \times 10^{-13}$.

\section{Results for Unrelated Data}


A second data set consists of exome sequence calls, blood pressure phenotypes at a single time point, and simulated phenotypes on a large set of unrelated individuals. Like the family data set, these individuals are Mexican Americans; however, they were independently ascertained and do not overlap with the family data set. 

\subsection{Size and power study using simulated traits (SIMPHEN.1--200)}
\label{sec:unrel-size-power}

200 simulation replicates of the trait SBPs and DBP are provided. However, GAW19 organizers did not distribute the exact simulation details, except stating that ``The set of causal variants is somewhat different since this is exome data rather than the full sequence data that was provided last time, and so not all of the GAW18 variants, regulatory ones in particular, are present in the new dataset." This precludes a precise size and power study. 

For ease of comparison, we test the same 6 variants displayed in Table~\ref{table:map4-pvalues} (for family data) against the bivariate trait (SBP,DBP) for all 200 simulation replicates and report the rejection rates in Table~\ref{table:unrel-map4-pvalues}. In the model we include covariates {\tt sex}, {\tt age}, {\tt BPMed}, {\tt Smoke}, and their pairwise interactions and use the SNP-based genetic relation matrix for modeling additive polygenic inheritance.

\begin{table}
\centering
{\footnotesize
\begin{tabular}{llcc}
\toprule
SNP Name in Family Data & SNP Name in Unrelated Data & MAF in Unrelated Data & Rejection Rate \\
\midrule
3-47913455 & not found   & \\
3-47956424 & rs1137524   & 0.3435 & 1.00 (0.00) \\
3-47957741 & rs138046751 & 0.0005 & 0.09 (0.02) \\
3-47957996 & rs2230169   & 0.0229 & 1.00 (0.00) \\
3-48040283 & rs11711953  & 0.0281 & 1.00 (0.00) \\
3-48040284 & var\_3\_48040284 & 0.0070 & 0.12 (0.02) \\
\bottomrule
\end{tabular}
}
\caption{Empirical rejection rates (standard errors in parenthesis) for testing 5 variants in the MAP4 gene against the bivariate (SBP,DBP) trait, based on simulation data in files SIMPHEN.1-SIMPHEN.200 for 1943 unrelated individuals. The first two columns contrast the SNP names of the same variants in the family and unrelated data respectively.}
\label{table:unrel-map4-pvalues}
\end{table}

\subsection{QTL analysis of the real, bivariate phenotypes (DBP and SBP)}
\label{sec:runrel-real-dbp}

The phenotypes SBP and DBP measured at the first examination are available for 1943 unrelated American Mexicans. We analyzed all SNPs and bivariate traits (SBP, DBP). To read in all the data and run standard QC procedures took 1 minute and 16 seconds. QC excluded 10,191 SNPs and 93 individuals based on genotyping success rates below 98\%. The remaining 1,701,575 SNPs and 1,850 individuals are analyzed. The subsequent ped-GWAS analysis ran in 37 minutes and 5 seconds and included all of the results plotted in Table~\ref{table:unrel-realpheno-DBP-SBP}.  Since we exclude rare variants with MAF $\le 0.01$ in all individuals, p-values were calculated for 52,314 SNPs. Accordingly, the genome-wide significance threshold is $9.56 \times 10^{-7}$ or 6.02 on the $\log_{10}$ scale.

Estimated environmental effects and their interactions and variance components under the null model (no SNPs included) are listed in the top panel of Table~\ref{table:unrel-realpheno-DBP-SBP}. The bottom panel displays the Manhattan and QQ plots. The genomic inflation factor of 1.001 indicates no systematic bias. No SNPs pass the genome-wide significance level or FDR 0.05 threshold.

\begin{table}[ht!]
\centering
{\footnotesize
  \begin{tabular}{@{} lrr @{}}
    \toprule
Mean effects     & SBP & DBP \\ 
\midrule
$\mu$ & 94.87 (1.62) & 78.46 (0.95) \\
$\beta_{\text{Sex}}$ & 10.90 (1.63) & 4.62 (0.95) \\ 
$\beta_{\text{Age}}$ & 0.43 (0.05) & -0.13 (0.03) \\ 
$\beta_{\text{Sex} \times \text{Age}}$ & 0.38 (0.06) & 0.08 (0.04) \\ 
\midrule
Var. comp. & $\bSigma_a$ = $\begin{pmatrix} 43.15 & 17.03 \\ 17.03 & 12.07 \end{pmatrix}$ & $\bSigma_e$ = $\begin{pmatrix} 294.88 & 113.90 \\ 113.90 & 102.61 \end{pmatrix}$ \\ 
\bottomrule
  \end{tabular}
}
$$
\begin{array}{m{3.25in}m{3.25in}}
\includegraphics[width=3.25in]{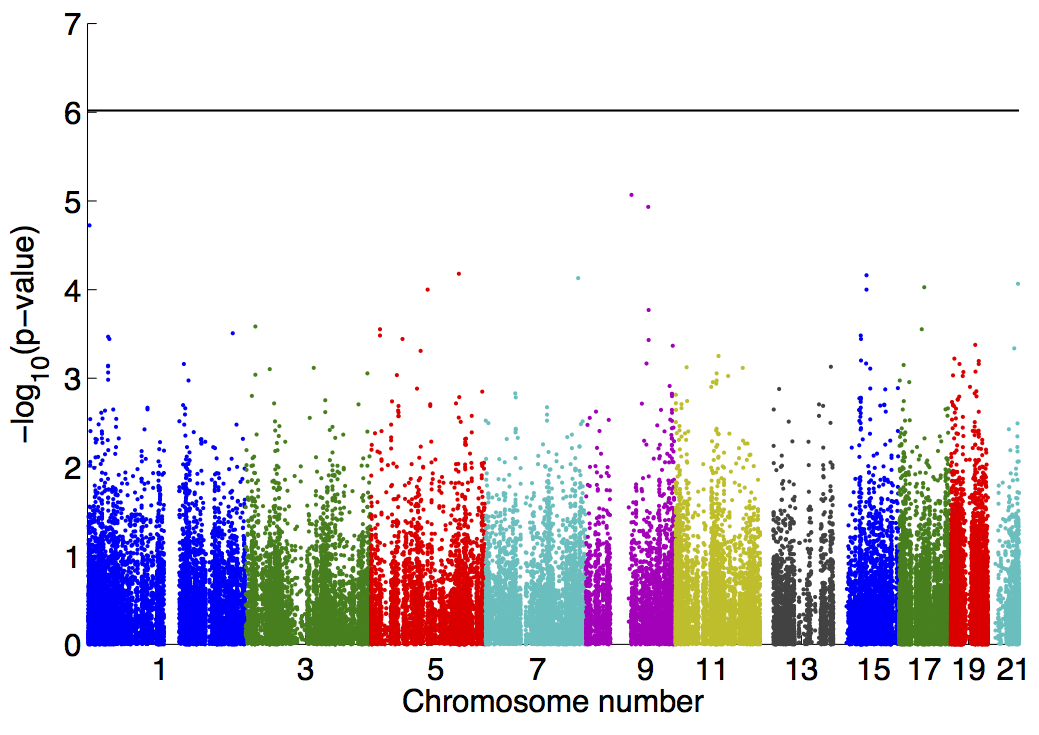} & \includegraphics[width=2.75in]{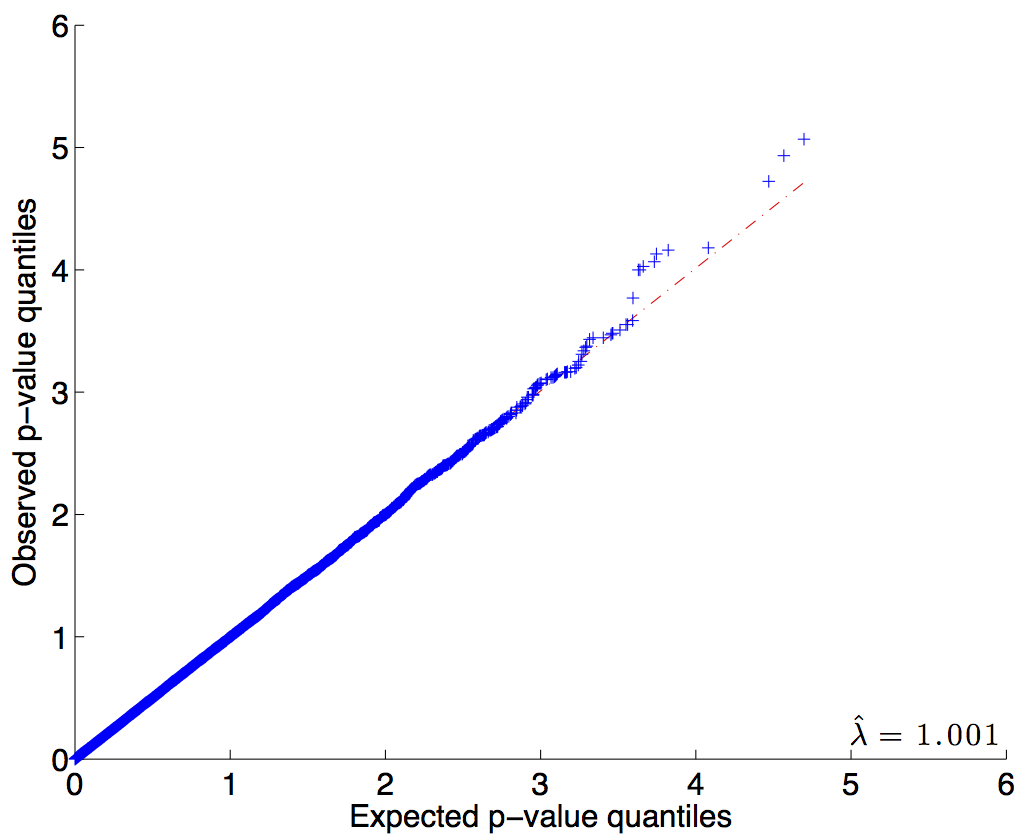} \\
\end{array}
$$
\caption{QTL analysis of the real, bivariate (SBP, DBP) trait for 1850 unrelated individuals and 52,314 SNPs with MAF $> 0.01$.
Top: Mean effects (standard errors in parenthesis) and variance components under the null model using GRM with all individuals. Bottom: Manhattan plot (left) and QQ plot (right). The horizontal line represents the genome-wide significance level; no SNPs pass this level. Total run time on a laptop with Intel Core i7 2.6 GHz CPU and 16 GB RAM is {\bf 39 minutes}.}
\label{table:unrel-realpheno-DBP-SBP}
\end{table}


\section{Conclusions}


All analyses in this article use \Mendel\ v14.3, which is freely available at \url{www.genetics.ucla.edu/software}. Pedigree GWAS (Option 29) in \mendel\ proves to be an extremely efficient and versatile implementation for large-scale QTL analysis. Most competing programs ignore multivariate traits and outliers altogether. See \citep{Zhou13PedScore} for a side-by-side comparison with the {\sc FaST-LMM} program. Here we have emphasized \mendel's flexibility in specifying the global kinship matrix, adjusting for confounding, and capturing interactions. These assets, plus its raw speed, make it an ideal environment for QTL mapping.  \mendel\ continues to mature, and geneticists are advised to give it a second look for genetic analysis \citep{Lange13Mendel}. 

\section*{Acknowledgments}

The authors gratefully acknowledge the NIH grants GM053275 and HG006139 and the NSF grant DMS-1310319.


\bibliography{gaw19_notes}

\begin{thebibliography}{}

\bibitem[Day-Williams et~al., 2011]{Day-Williams11LinkageWoPedigree}
Day-Williams, A.~G., Blangero, J., Dyer, T.~D., Lange, K., and Sobel, E.~M.
  (2011).
\newblock Linkage analysis without defined pedigrees.
\newblock {\em Genetic Epidemiology}, 35(5):360--370.

\bibitem[Lange, 2002]{Lange02GeneticsBook}
Lange, K. (2002).
\newblock {\em Mathematical and Statistical Methods for Genetic Analysis}.
\newblock Springer-Verlag, New York, second edition.

\bibitem[Lange et~al., 2014]{Lange14NextGenStatGene}
Lange, K., Papp, J., Sinsheimer, J., and Sobel, E. (2014).
\newblock Next-generation statistical genetics: Modeling, penalization, and
  optimization in high-dimensional data.
\newblock {\em Annual Review of Statistics and Its Application}, 1(1):279--300.

\bibitem[Lange et~al., 2013]{Lange13Mendel}
Lange, K., Papp, J., Sinsheimer, J., Sripracha, R., Zhou, H., and Sobel, E.
  (2013).
\newblock Mendel: the swiss army knife of genetic analysis programs.
\newblock {\em Bioinformatics}, 29:1568--1570.

\bibitem[Lange et~al., 2005]{Lange05association}
Lange, K., Sinsheimer, J., and Sobel, E. (2005).
\newblock Association testing with {M}endel.
\newblock {\em Genetic Epidemiology}, 29:36--50.

\bibitem[Zhou et~al., 2014]{Zhou13PedScore}
Zhou, H., Blangero, J., Dyer, T.~D., Chan, K.-H., Sobel, E.~M., and Lange, K.
  (2014).
\newblock Fast genome-wide {QTL} association mapping on pedigree and population
  data.
\newblock {\em submitted}.

\end{thebibliography}
\bibliographystyle{apalike}

\end{document}